\documentclass[conference]{IEEEtran}
\IEEEoverridecommandlockouts
\usepackage{cite}
\usepackage{amsmath,amssymb,amsfonts}
\usepackage{algorithmic}
\usepackage{graphicx}
\usepackage{textcomp}
\usepackage{xcolor}
\usepackage{dblfloatfix}
\def\BibTeX{{\rm B\kern-.05em{\sc i\kern-.025em b}\kern-.08em
    T\kern-.1667em\lower.7ex\hbox{E}\kern-.125emX}}
\begin{document}

\makeatletter
\newcommand{\newlineauthors}{%
  \end{@IEEEauthorhalign}\hfill\mbox{}\par
  \mbox{}\hfill\begin{@IEEEauthorhalign}
}
\makeatother

\title{Gluing Neural Networks Symbolically Through Hyperdimensional Computing
{\footnotesize \textsuperscript{}}
\thanks{}
}

\author{\IEEEauthorblockN{Anonymous Authors}}

\author{\IEEEauthorblockN{Peter Sutor\textsuperscript{\textsection}}
\IEEEauthorblockA{\textit{Department of Computer Science} \\
\textit{University of Maryland}\\
College Park, Maryland, USA \\
psutor@umd.edu}
\and
\IEEEauthorblockN{Dehao Yuan\textsuperscript{\textsection *}}
\IEEEauthorblockA{\textit{Department of Computer Science} \\
\textit{University of Maryland}\\
College Park, Maryland, USA \\
dhyuan@umd.edu}
\and
\IEEEauthorblockN{Douglas Summers-Stay}
\IEEEauthorblockA{\textit{Comp. and Info. Sciences Directorate} \\
\textit{CCDC Army Research Laboratory}\\
Adelphi, Maryland, USA \\
douglas.a.summers-stay.civ@mail.mil}
\newlineauthors
\IEEEauthorblockN{Cornelia Ferm\"{u}ller}
\IEEEauthorblockA{\textit{Department of Computer Science} \\
\textit{University of Maryland}\\
College Park, Maryland, USA \\
fermulcm@umd.edu}
\and
\IEEEauthorblockN{Yiannis Aloimonos}
\IEEEauthorblockA{\textit{Department of Computer Science} \\
\textit{University of Maryland}\\
College Park, Maryland, USA \\
jyaloimo@umd.edu}\\
\newlineauthors
{\footnotesize \textsuperscript{*} Corresponding Author, \textsuperscript{\textsection} These authors contributed equally and are listed in alphabetical order}
}

\maketitle

\begin{abstract}
Hyperdimensional Computing affords simple, yet powerful operations to create long Hyperdimensional Vectors (hypervectors) that can efficiently encode information, be used for learning, and are dynamic enough to be modified on the fly. In this paper, we explore the notion of using binary hypervectors to directly encode the final, classifying output signals of neural networks in order to fuse differing networks together at the symbolic level. This allows multiple neural networks to work together to solve a problem, with little additional overhead. Output signals just before classification are encoded as hypervectors and bundled together through consensus summation to train a classification hypervector. This process can be performed iteratively and even on single neural networks by instead making a consensus of multiple classification hypervectors. We find that this outperforms the state of the art, or is on a par with it, while using very little overhead, as hypervector operations are extremely fast and efficient in comparison to the neural networks. This consensus process can learn online and even grow or lose models in real time. Hypervectors act as memories that can be stored, and even further bundled together over time, affording life long learning capabilities. Additionally, this consensus structure inherits the benefits of Hyperdimensional Computing, without sacrificing the performance of modern Machine Learning. This technique can be extrapolated to virtually any neural model, and requires little modification to employ - one simply requires recording the output signals of networks when presented with a testing example.
\end{abstract}

\begin{IEEEkeywords}
hyperdimensional, computing, symbolic, neural, bundling
\end{IEEEkeywords}

\section{Introduction}

Every year, many neural networks are trained to solve many different problems. However, with every new network trained that outperforms its predecessors significantly, previous networks are all but discarded. There is no reason their results can't be reused to further improve overall performance. We would like a solution that integrates models together to solve problems, despite differences in performance, network architecture, or even modality. Furthermore, it would be preferable to be able to grow such a consensus structure as more models are created, have the capability to update old models, and even remove models from the consensus. 

In this paper, we explore one possible solution that can afford us these properties, by using Hyperdimensional Computing (HDC) and Hyperdimensional Vectors (hypervectors) to build simple, but powerful models that can be attached as a post-processing step to virtually any neural network. As the network observes examples and makes predictions, these models learn along with it by directly encoding the output signals that form a prediction in the final layer of the network, and incorporating them into existing hypervector models. Through a special consensus summation, example encodings are accumulated into singular hypervectors that describe a class, which are then also accumulated into a single vector that can differentiate classes. Such models are then also accumulated across differing networks to form a single predictive hypervector that can predict the output class through an efficient consensus across all networks.

We demonstrate the functionality of these hypervectors by employing them on several data sets, to show how much the consensus improves the performance. Furthermore, we demonstrate how such consensus models can be dynamically changed over time. This includes some surprisingly flexible abilities, such as online learning in real time, training on the training set until 100\% accuracy is achieved through handling erroneous predictions, similarly incorporating the testing set to achieve 100\% accuracy, tolerance to neural network models not being available, forming ``memories" that can grow over time, the capability to dynamically add and remove neural network models, and the ability to learn during the training of a neural network, where hypervectors tend to quickly reach near optimal performance as they are better at few-shot learning. Usage of hypervector consensus is a post-processing step, meaning pre-trained networks can be used by simply testing them across the training data set, recording the hypervector encodings of their output signals, and then incorporating those into the consensus model. The technique is agnostic to the architectures of networks used or their modalities. 

The rest of this paper proceeds as follows. In section \ref{background}, we give a brief review of necessary background information on Hyperdimensional Computing with hypervectors. In section \ref{related}, we discuss prior work related to this paper and other existing relevant methods that achieve similar goals. In section \ref{hdglue}, the algorithm itself for gluing neural models together is described in detail and discussed. In section \ref{methods}, the methodology used to test our performance is described, including what data sets were used and how tests were set up. In section \ref{results}, we present our results on tests and discuss their implications. In section \ref{discussion}, we analyze our results to suggest the pros/cons of our gluing technique and future work to pursue. Finally, section \ref{conclusion} concludes the paper, drawing broad results, while section \ref{acknowledgements} acknowledges those who made this work possible.

\section{Background}\label{background}

Here, we give a very brief review of some of the relevant background on Hyperdimensional Computing with binary vectors. Much of this is based on Pentti Kanerva's work \cite{kanerva2009hyperdimensional}. Binary hypervectors are traditional binary vectors, but with a component length that approaches hyperdimensional proportions, typically in the thousands. As the number of possible binary vectors of length $n$ increase exponentially as $n$ increases, the $2^n$ hypervectors in the vectorspace quickly approach massive values. At hyperdimensional lengths, a randomly selected hypervector is almost guaranteed to be near orthogonal to another random hypervector. In terms of Hamming distance, this means a distance of about half the length of the hypervectors, as every bit is effectively randomly selected. This property can be abused to detect correlation, by finding hypervectors that are significantly closer, or clustered. At lengths of thousands of bits, a deviance in expected Hamming distance of even a few hundred bits correspond to many standard deviations of significance. Hypervectors that deviate by only, say, one thousand bits out of ten thousands are considered to basically be the same, ``noisy" hypervectors. Thus, Hyperdimensional Computing is great at detecting similarity, even in noisy settings, assuming the hypervectors are meaningfully constructed.

The most basic operation is the element-wise Exclusive-Or (XOR). The XOR operation's output is the hypervector of differences, where a 1 means the vectors differed and a 0 means they did not. In this paper, element-wise XOR is represented as the $\oplus$ operator. Next, is the permutation of bits, where a hypervector's bit locations are permuted according to a provided sequence re-ordering, represented as $\Pi$. Thus, the term $\Pi X$ represents a (possibly random) permutation of hypervector $X$'s bits. Repeating the permutation $n$ times is represented as a power, such as $\Pi^n X$. Lastly, but most importantly, is the consensus summation. When $n$ hypervector terms are consensus summed together, the resulting hypervector is constructed of the most popular bit across all the terms in each component, so a bit-wise democratic vote. Consensus sum is expensive to recompute, as adding another term requires re-computing all the additions, or remembering a long vector of integers. Unless otherwise stated, the $+$ operator, when applied to Hypervectors, is assumed to be consensus summation.

For meaningful constructions of hypervectors, ``bundling" - an aggregation of hypervectors - and ``binding" - an association between two or more hypervectors - are typically used. For example, a set can be represented as the XOR bundling of all hypervectors in the set. XOR-ing the result with a hypervector that is already in the set effectively removes it, as $X \oplus X = \hat{0}$, and $Y \oplus \hat{0} = Y$. Ordered sequences can be achieved through a permute-multiply paradigm, thus the sequence $abcd$ can be represented as a hypervector using their hypervector counterparts $A$, $B$, $C$, and $D$ as: $A \oplus \Pi B \oplus \Pi^2 C \oplus \Pi^3 D$. A new element in the sequence can be added by first permuting with $\Pi$ once more, then XOR-ing with the hypervector representing the element. The reverse removes an element. Any subsequence can be removed or skipped, similarly. Finally, ``data records" can be achieved using an add-multiply paradigm. For example, if we have a database where each entry is of the form [name, sex, age], then [Jim, Male, 25] would be encoded as: $J \oplus N + M \oplus S + X \oplus A$, where $N$, $S$, and $A$ represent the name, sex and age records symbolically, and $J$, $M$, and $X$ are hypervector encodings of these attributes.

More generally, given basis ${F_1, F_2, \dots, F_n}$ and corresponding data ${D_1, D_2, \dots, D_n}$, a record can be constructed as:

\begin{equation}\label{eq:records}
    R = D_1 \oplus F_1 + D_2 \oplus F_2 + \dots + D_n \oplus F_n
\end{equation}

\noindent To recall a piece of data, say, $D_i$, one can simply compute:

\begin{equation}\label{eq:probing}
    R \oplus D_i = F_i + \eta \approx F_i
\end{equation}

\noindent where $\eta$ is a ``noise" vector that arises from consensus. By cycling through possible $F_j$, the correct field can be identified by finding the one with minimum Hamming distance. If needed, we can adjust \eqref{eq:records} to account for weighted terms. Simply introduce weights $a_1, a_2, \dots, a_n$ such that $a_1 + a_2 + \dots + a_n = n$, for $n$ terms being added. Instead of each term contributing one vote for a particular bit, it now counts as a fractional vote:

\begin{equation}\label{eq:recordsweighted}
    R = a_1 D_1 \oplus F_1 + a_2 D_2 \oplus F_2 + \dots + a_n D_n \oplus F_n
\end{equation}

\begin{figure*}[t]
\centerline{\includegraphics[width=0.99\textwidth]{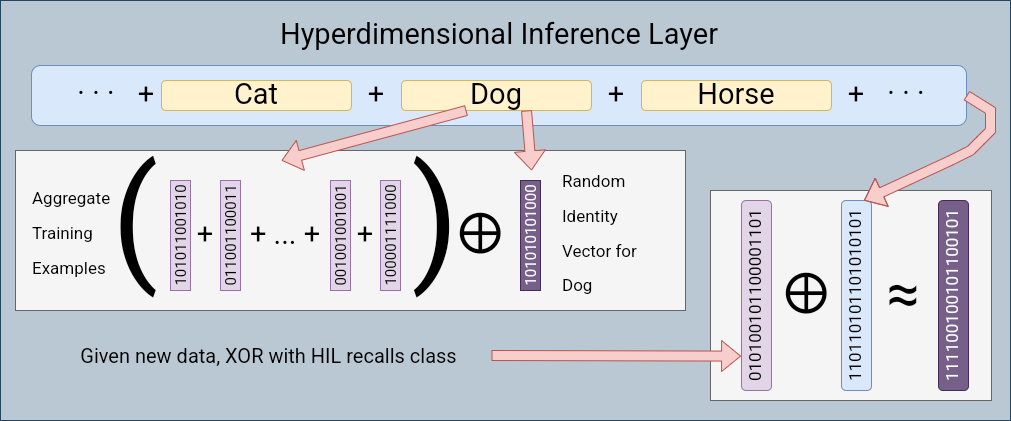}}
\caption{Visualization of how the Hyperdimensional Inference Layer (HIL) \cite{mitrokhin2019learning, mitrokhin2020symbolic} learns to classify. Given a training set of hypervectors encoding features for an animal, each class bundles the associated training data and binds it to a random, symbolic representation of the class. Then, we bundle across all classes. To recall a class, simply XOR a new feature vector with the HIL, the closest class hypervector will be the model prediction.}
\label{fig:hilprocess}
\end{figure*}

Hypervectors in consensus exist in a state of pseudo-superposition, such that the XOR will collapse incorrect matches into random noise, as they will be near orthogonal in hyperdimensional spaces. Only the correct match will stand out with statistical significance from the noise, perturbing the noise vector to be non-random. Likewise, $D_i$ can be found by probing with $F_i$. The technique in \eqref{eq:records} and \eqref{eq:probing} is key to our consensus strategy of fusing networks. Consider $F_1 = \texttt{IF}$ and $F_2 = \texttt{THEN}$ - then $D_1 \oplus F_1 + D_2 \oplus F_2$ becomes a simple if-then inference structure, where given appropriate $D_1$, the XOR with $D_1$ can indicate whether the IF condition has been satisfied, from which a probe for the THEN post-condition reveals what we should do through $D_2$. One can easily see that such inference structures can be used to solve problems - we will refer to such usage of hypervectors as Hyperdimensional Inference (HI). Relevant to this paper, a structure where many $F_i$ are presented, and we must choose the correct one given a data point $D$, will be referred to as a Hyperdimensional Inference Layer (HIL), visualized in Fig.~\ref{fig:hilprocess}.

\section{Related Works}\label{related}

In this section, we will discuss relevant works to the consensus technique presented in this work. This is broadly split into two subsections; prior work that is used or is relevant, and works that solve similar problems.

\subsection{Prior Work}

This work is an extension of a series of works. First, \cite{widdows2015reasoning, rahimi2016hyperdimensional, rahimi2018efficient, rachkovskiy2005sparse, sutor2018computational} describe some methods of encoding arbitrary data into functional hypervectors. Namely, ideas presented in \cite{rachkovskiy2005sparse} and \cite{sutor2018computational} were used to facilitate the process of converting output signals from networks into hypervectors. This work builds on Hyperdimensional Active Perception \cite{mitrokhin2019learning} and the notion of forming a memory unit from the class space much like in \eqref{eq:records} and probing this with \eqref{eq:probing} to discover the best classification. More specifically, a followup paper shows this concept applied to the standard problem of classifying images, where neural network outputs were being directly encoded into memories \cite{mitrokhin2020symbolic}. Here, the authors use neural hashing networks to convert images into short, rankable, similarity-based binary vectors. These vectors were then projected into hyperdimensional spaces, and then used similarly to what is described in section \ref{hdglue} to fuse architecturally different hashing network outputs into a more powerful, consensus model. Our work is most similar to this, but with the novel additions of extremely broadening the scope of networks that can be fused, directly encoding output signals and not the output, and presenting life-long learning strategies that can increase the performance.

\subsection{Similar Work}

As stated before, our work is an advancement of \cite{mitrokhin2020symbolic}. However, there are other works that attempt to either fuse data with hypervectors or encode neural networks in hypervectors. For example, in \cite{benediktsson1997classification}, hyperdimensional data is fused together to improve classification accuracy by incorporating many modalities of data. Effectively, we use a similar idea in our work, except the data is the direct encoding of neural output signals. A followup work to this uses a neural network to aid the consensus classification from hypervectors as a second stage of processing \cite{benediktsson1999classification}. In theory, this is something that can be used in place of the method of determining the correct class presented in section \ref{hdglue}. Furthermore, the idea of using superpositions to compress aspects of networks into one is explored in \cite{cheung2019superposition}, where the idea is applied to network parameters to fuse them into one. We focus on the output signals of only a part of the network. Finally, the problem of classification using hypervectors has been explored in previous works, such as \cite{rahimi2018efficient} and \cite{kleyko2018classification}, in which the basics of record-based classification ala \eqref{eq:records} and \eqref{eq:probing} were explored, but do not attempt data fusion, consensus or encoding of network signals, but focus on the task of classifying itself.

Finally, it is necessary to mention prior work that does not directly use HD techniques or hypervectors, but still enables ensembling or consensus among different models. These are mentioned mainly to show how our work differs from these. Ensemble computing/learning generally makes use of multiple learning algorithms to improve the overall predictive performance that can be achieved by any one model alone. A survey of popular techniques can be found in \cite{sagi2018ensemble, dietterich2002ensemble, polikar2012ensemble, dong2020survey}. The primary difference between our work here and work in ensembling is our goal to ``normalize" the space on which our model's predictions are aggregated, by using the same-size, binary hypervectors. In consensus learning, such as some of the techniques in \cite{dong2020survey}, the ensembling occurs through consensus from all models. Thus, all models participate in ``voting" on the final prediction. Our work can be viewed as ensemble consensus, but it should be noted that binary hypervectors and how they correlate spatially can have non-linear relationships. It's entirely possible to aggregate multiple models, but only receive a statistically significant response, given some input, for a few models, or perhaps only one. This arises due to the ``noisiness", present in hypervectors.

\section{The Hyperdimensional Gluing Algorithm}\label{hdglue}

In this section, we describe our novel algorithm for fusing the output signals of neural networks together to form a consensus structure for improved performance, which we refer to as Hyperdimensional Gluing, or HD-Glue, for short.

\subsection{Encoding Neural Output Signals as Hypervectors}\label{encodeweights}

The key cornerstone to being able to combine neural networks and hypervectors is the observation that a method of translating neural network outputs to hypervectors must be designed. In previous work, such as \cite{joshi2016language, imani2019quanthd, mitrokhin2019learning, mitrokhin2020symbolic, sutor2018computational}, the direct output of the network was used to do this. In other works, symbolic representations of output classes or values were used. However, these techniques are not necessarily extendable to all networks, or they do not sufficiently capture what the network has learned from the data, apart from broad classes. We inevitably reach the conclusion that parts of the network must be directly encoded. For the purpose of demonstration in this paper, we assume a neural network consists of multiple layers, where the final layer's output signals are combined to determine the output class. Suppose these signals are available; we can encode the neural output signals directly into hypervectors to get a more informed hypervector encoding that conveys ``why" the network made that classification choice.

To achieve conversion from neural signals to hypervector encodings, we further assume the signals are simple, real values. As such, we organize them into a real vector, sometimes referred to as an embedding. Embeddings are high level spaces that are projected into low level dimensions, thus capturing semantics of the high level input space by moving related objects in that space closer together. To convert a real vector embedding into a hypervector encoding, we recognize that the real vector itself is an ordered sequence of real values. We have existing hypervector encoding techniques to encode ordered sequences of numbers, such as \cite{joshi2016language}, but encoding real numbers themselves is more challenging. Fortunately, neural output signals can be normalized to a range of $[-1, 1]$ using the $\tanh$ function, which allows us to encode signals by using a sufficiently fine binning of values, and assigning each bin a hypervector. The techniques described in \cite{joshi2016language, imani2019quanthd, mitrokhin2020symbolic, sutor2018computational} and, namely, in \cite{mitrokhin2019learning}, can be used to create meaningful bin hypervectors by slowly changing the hypervector for -1 into the hypervector for 1 with each additional bin, by randomly changing a constant number of unchanged bits. Then, \eqref{eq:records} indicates that a real vector can be represented as a bundling of component positions (their symbolic hypervectors forming a basis), each one a binding of the signal value hypervector to the component ID (a randomly selected vector). The pipeline from input to hypervector encoding is shown in Fig.~\ref{fig:embedconvert}.

\begin{figure*}[t]
\centerline{\includegraphics[width=0.99\textwidth]{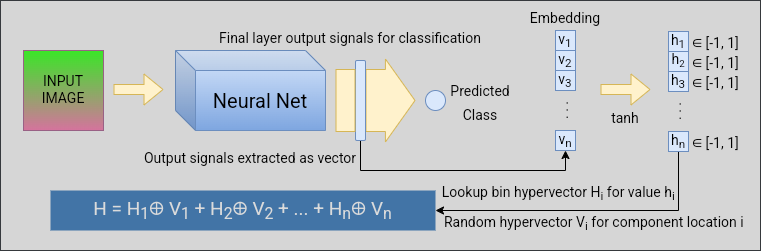}}
\caption{The pipeline for input to hypervector encoding of network features. First, an input is presented to a network, which generates output signals that predict the final class. These are copied as an embedding and exported from the network. Then, \texttt{tanh} is applied to each component value. The results are binned into hypervectors that represent the range [-1, 1], quantizing the values. Finally, we use random symbolic hypervectors to represent component positions and formulate the final encoding with \eqref{eq:records}.}
\label{fig:embedconvert}
\end{figure*}

\subsection{Training Hypervector Models for a Neural Network}\label{singlenet}

Suppose we only have one neural network to start with. Training examples are presented to the network, which prompt activations in the final layer that are used to form the prediction for the network, outputting signals for each neuron. Each output is encoded by quantizing it as the nearest bin hypervector in the range $[-1, 1]$. Then, a consistent list of hypervectors representing component positions binds the signal hypervector to the appropriate bin. The aggregation of these hypervectors across all components through consensus summation forms the final hypervector representation of the real vector of signals. This can now be stored as a ``memory". For each new training example for a particular class, memories are aggregated by class from the gold standard labels - these are consensus summed once every training example for the class has been seen. Consensus sum forces an ``average" vector to be created that is forced to throw out the unpopular vote in every component and choose the most commonly represented bit across the terms. This facilitates ``learning". Thus, the bundled memories for each class are a learned representation.

Classes are represented by random hypervectors themselves as ID hypervectors. Once training is complete, each learned hypervector corresponding to a class is bound to the corresponding hypervector representation. Finally, these terms are themselves bundled, forming a final classification hypervector. When a new example is provided at testing time, the neural network creates a new real vector of signals, which is quantized to a hypervector encoding - to identify the correct class we XOR this with the model and probe for the correct class as shown in \eqref{eq:probing}. Once again, this process is visualized in Fig.~\ref{fig:hilprocess}. If training is resumed, hypervector models can easily incorporate new training examples into the aggregates, further learning. With a fine enough bin, virtually no information is lost from real vectors to hypervectors. As we will see in the results in section \ref{results}, this transformation does not impact the performance negatively with sufficiently hyperdimensional hypervector lengths.

\subsection{Forming Multiple Trained Models for a Neural Network}\label{multihypervectors}

Since hypervectors exist as discrete entities, they can be used either as a form of memory, or actual models that solve problems. As such, there is no reason multiple models cannot be used. In this context, given a neural network, one can have many hypervectors that see a subset of the training data, then predict the result at testing time in tandem. Consensus can be achieved by repeating the same records strategy in \eqref{eq:records}. Each hypervector model can be assigned its own random ID hypervector, which is bound to the model. These bound terms are then bundled with consensus summation. Given a new example, we can probe the singular model to determine which class has the most agreement across models.

This strategy can be used in other interesting ways. Suppose we train a hypervector model from a neural network to solve a problem. If it doesn't achieve 100\% accuracy, we can collect the examples that were not correctly classified, and then form a new hypervector model to address these partially. If we still do not achieve 100\% accuracy, simply repeat the process until no improvement can be made. The weighted consensus summation in \eqref{eq:recordsweighted} can be used to account for each subsequent model observing less and less training examples, where the weights are the proportion of the training examples used for that model. In Fig.~\ref{fig:errorcorrecting}, this process is demonstrated with semi-realistic numbers. The model $A$ is trained, but only achieves 76\% accuracy, thus it contributes 0.76 to the final model. Then, model $B$ handles the remaining 24\%, but only achieves 70\% accuracy there, thus it contributes $0.24 \times 0.70 = 0.168$ to the final model. Updating $A$ to the weighted average of $A$ and $B$ improves its performance. A fleet of such error corrected models can be used in consensus to effectively reach 100\% accuracy, if desired. In an online learning setting, we can grow more and more models to handle errors. If we have too many models, simply consensus the worst performing models into a single hypervector, and throw away the individual models. This allows life long learning and improvement.

\begin{figure}[t]
\centerline{\includegraphics[width=0.475\textwidth]{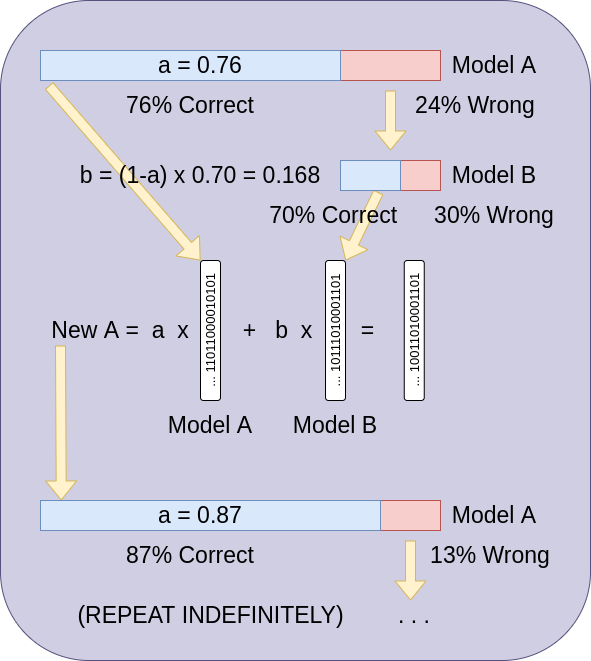}}
\caption{Example of error correction, where the aggregated hypervector model achieves 100\% accuracy on the data set. As model A is trained, the examples it got wrong are collected and used to train model B. This process can repeat until no improvement can be made. A weighted consensus sum is used to prevent over correction. A fleet of error corrected models can be trained to achieve 100\% accuracy on the training set.}
\label{fig:errorcorrecting}
\end{figure}

\subsection{Forming Trained Models for Multiple Neural Networks}\label{multinets}

Now that we've established the training process for a single neural network's output signals, we can address doing this for many neural networks to fuse together their output signals symbolically into a single consensus model. We note that subsections \ref{encodeweights}, \ref{singlenet}, and \ref{multihypervectors} set up the pre-processing steps up until this point. To fuse the models, we repeat the same strategy in subsection \ref{multihypervectors}. A record is created using \eqref{eq:records}, where random ID hypervectors are generated for each neural network and bound to the hypervector models. These terms are then consensus summed. The result is a single predictive hypervector model that takes into account all neural networks and lets the ``vote" on the correct class when given a new testing example. Thus, ``gluing" the models together through a symbolic hyperdimensional representation, a process that we will refer to a HD-Glue for short.

Interestingly, as is the case with all record based formulations, you do not need to incorporate every network in the voting process. It suffices that one network votes, collapsing the contributions of others into noise. However, at that point, it would be better to use the network's learned hypervector model. This is more useful for when some networks aren't available as a resource. Furthermore, multiple HD-Glue models can be formed, in a process like the one described in subsection \ref{multihypervectors}, to ensure that the consensus of networks achieves 100\% accuracy on the training set.

\section{Methodology}\label{methods}

In this section, our methodology for testing the HD-Glue technique is described, including what data sets we use for testing, what sort of tests were set up and their purpose, and what a successful result looks like.

\subsection{MNIST Data-set}\label{MNIST}

Here we describe our process for testing HD-Glue on the MNIST data-set \cite{deng2012mnist}. The MNIST data set contains 60k examples of hand-written digits 0-9 that are compressed into 40-by-40 pixel greyscale images. It is a common benchmark for new learning techniques.

\subsubsection{Purpose}

The purpose of testing on this data-set is to benchmark the HD-Glue technique against other common algorithms that can use real vectors to fuse the outputs of neural networks. These are simple, comparable methods to HD-Glue, that can be dynamically constructed. It should be noted, however, that these methods do not work on vectors of different sizes, while HD-Glue is agnostic to what the vector looks like as it is a purely symbolic construction.

\subsubsection{Image Encoder Networks} 

In order to handle the issue of needing a consistent number of output signals for comparing to other methods, we use simple image encoder networks that learn representations for the digits, with a final layer used for classification - the signals that feed into this are kept at consistent sizes.

\subsubsection{Testing Procedure}

We train a fixed number of encoder networks with different parameters. These are then cached and used with our benchmark methods and HD-Glue. For HD-Glue, each image generates a hypervector encoding, which are used with the techniques described in section \ref{hdglue}. Every step in section \ref{hdglue} is tested for performance, to observe how the overall performance of HD glue compares to its constituents. As these methods can learn in online settings, we also graph the performance between few shot examples, to the full data-set, across all methods. It should be noted, that for MNIST we generate 5 classes of models that only discriminate 2 digits at a time. This differs in further tests, but demonstrates HD-Glue's capabilities when the classes are not consistent across models. Additionally, we also use these to demonstrate HD-Glue in an online learning setting, where new classes appear after a certain amount of examples are seen, such that an unbalanced number of training examples are seen per digit. 

\subsection{CIFAR-10}\label{cifar10}

Here we describe testing HD-Glue on the CIFAR-10 data-set \cite{krizhevsky2009learning}. CIFAR-10 contains 10 different classes of objects/animals, with 60k 32-by-32 pixel color images.

\subsubsection{Purpose}

CIFAR-10 is a more rigorous benchmark, with many diverse pre-existing models to use for consensus, with output signals easily available. Thus, in addition to our encoder tests, common models exist for use in tests, to allow diverse networks and different embedding lengths.

\subsubsection{Pre-Existing Models Used}

VGG11 \cite{simonyan2014very} models were used to test against other benchmark methods, each trained with different initializations. This way, embedding lengths across models acting in consensus can remain constant, otherwise benchmark methods could not be used - only HD-Glue can handle this.

\subsubsection{Testing Procedure}

This does not differ much from MNIST's procedure. The only difference is that all models used have the same number of classes. Similar to MNIST, we test against other benchmark methods, but we also test HD-gluing many differing models together. As shown in \cite{mitrokhin2020symbolic}, this can greatly boost performance as the consensus process can select the model that best solves a particular training example, so even if a weaker model outperforms a stronger model on very few examples, those examples will be selected when appropriate.

\subsection{CIFAR-100}\label{cifar100}

The CIFAR-100 data-set \cite{krizhevsky2009learning} differs only in number of training classes, ten times that of CIFAR-10, and has much fewer training examples per class, making it a more difficult problem to solve. This is a very rigorous benchmark of the HD-Glue method, with the most room to improve consensus performance over constituent models. Our methodology here is the same as CIFAR-10 in subsection \ref{cifar10}. We differ only in testing the benchmarks across different architectures, and then in a separate experiment testing differing embedding lengths, to demonstrate that HD-Glue is agnostic to the architectures of the neural networks. Pre-existing models used were VGG11, VGG13, VGG16, and VGG19 from the VGG family \cite{simonyan2014very}, and ResNet18 and ResNet34 from the ResNet family \cite{he2016deep}.

\begin{table*}[hbt!]
\caption{HD-Glue performance vs. benchmark methods on the MNIST data-set}
\begin{center}
\begin{tabular}{|c|c|c|c|c|c|c|c|c|c|}
\hline
\multicolumn{10}{|c|}{\textbf{MNIST Data-set Benchmarking Results}} \\
\hline
\textbf{\textit{Size}}& \textbf{\textit{KNN}}& \textbf{\textit{Linear SVM}}& \textbf{\textit{RBF-SVM}}& \textbf{\textit{DT}}& \textbf{\textit{RF}}& \textbf{\textit{Log-Reg}}& \textbf{\textit{AdaBoost}}& \textbf{\textit{Naive Bayes}}& \textbf{\textit{HD-Glue}}  \\
\hline
10& 33.0\%& 50.8\%& 50.8\%& 29.5\%& 43.2\%& 51.9\%& 28.4\%& 50.8\%& \textbf{56.3\% (+4.4\%)} \\
\hline
50& 53.9\%& 48.4\%& 48.6\%& 41.9\%& 45.1\%& 58.1\%& 24.1\%& 49.7\%& \textbf{63.1\% (+5.0\%)} \\
\hline
100& 54.0\%& 49.2\%& 54.7\%& 50.0\%& 55.2\%& 63.9\%& 27.7\%& 55.0\%& \textbf{64.5\% (+0.6\%)} \\
\hline
500& 66.3\%& 61.3\%& 70.3\%& 61.1\%& 62.8\%& 73.4\%& 22.1\%& 63.6\%& \textbf{74.2\% (+0.8\%)} \\
\hline
1000& 69.5\%& 63.8\%& 73.6\%& 62.1\%& 63.5\%& 74.6\%& 35.5\%& 64.2\%& \textbf{75.2\% (+0.6\%)} \\
\hline
\end{tabular}
\label{tab:mnist}
\end{center}
Results for MNIST are shown above on our benchmarks. Here, five image encoder networks were used, each trained to only handle two of the ten MNIST digits. The size column is the number of training examples shown to the neural network, thus the number of embeddings generated in the output layer. The column for Log-Reg is the logistic regression of the element-wise average of the image encoder embeddings, while HD-Glue is our algorithm from the encoded hypervectors of the networks. The remaining columns are other common algorithms that run directly on the real vector embeddings, by performing an element-wise average across encoder embeddings. These include KNN ($k$ Nearest Neighbors \cite{fix1989discriminatory}), Linear SVM \cite{cortes1995support}, RBF-SVM (Radial Basis Function SVM) \cite{cortes1995support}, DT (Decision Tree \cite{quinlan1986induction}), RF (Random Forest \cite{ho1995random}), AdaBoost \cite{freund1997decision}, and Naive Bayes \cite{rish2001empirical}. The best performing model is bolded and states how much better it is than the second best model in terms of percentages. As we can see, HD-Glue outperforms all, especially in few shot settings - meaning HD-Glue can be used to boostrap a consensus model before the networks are even done training. This aligns with the results in \cite{mitrokhin2020symbolic}.
\end{table*}

\section{Results}\label{results}

In this section we present our results on the data-sets in section \ref{methods}, as per the methodology described.

\subsection{MNIST Results}\label{mnistres}

Our results are shown in Table~\ref{tab:mnist}. In this experiment, 5 encoders were created, used to discriminate between 2 of the MNIST digits, each. Partitioning classifiers like this simulates what would happen in an online setting where new classes and new networks simultaneously come into existence. Each one is steadily presented more and more examples, tracking the accuracy from few-shot settings to large numbers of examples. By keeping the output layers of encoder networks the same number of neurons, other benchmarking algorithms can be used to compare to HD-Glue.

Furthermore, the results for our online learning experiment on MNIST are shown in Table~\ref{tab:mnistonline}. The results here do \textbf{not} use any error correction nor multiple consensus models per class. The performance is broken down on a class-by-class basis, tracking the performance as each class is presented new examples in an online fashion. Once every existing class has seen 100 new examples, 2 new classes are added by adding a new model, and 100 examples are presented again for each class. This simulates online learning across an unbalanced dataset. By the end, 3000 examples have been presented, the final performance being the same as if training on all 3000 examples at once. We can see that HD-Glue gracefully handles new classes and quickly learns them. Despite digits 2 and 3 seeing 400 examples, and digits 8 and 9 seeing only 100, the latter outperforms the former. This likely reflects the difficulty of the task, moreso than the amount of data. Existing classes gracefully decay in performance as new classes are introduced, as the single hypervector predictive model is forced to discriminate between more and more classes.

\begin{table}[hbt!]
\caption{HD-Glue online learning}
\begin{center}
\begin{tabular}{|c|c|c|c|c|c|}
\hline
\multicolumn{6}{|c|}{\textbf{MNIST Online Learning Results}} \\
\hline
 & \multicolumn{5}{|c|}{\textbf{Number of classes with 100 new examples each}} \\
\hline
\textbf{Class}&
$\mathbf{2 \times 100}$&
$\mathbf{4 \times 100}$& $\mathbf{6 \times 100}$& 
$\mathbf{8 \times 100}$& $\mathbf{10 \times 100}$ \\
\hline
Digit 0& 99.9\%& 99.3\%& 98.1\%& 94.1\%& 94.3\% \\
\hline
Digit 1& 100.0\%& 98.1\%& 97.6\%& 98.2\%& 97.9\% \\
\hline
Digit 2& & 86.9\%& 90.8\%& 86.7\%& 82.9\% \\
\hline
Digit 3& & 98.9\%& 89.2\%& 82.4\%& 80.0\% \\
\hline
Digit 4& & & 98.0\%& 95.2\%& 84.4\% \\
\hline
Digit 5& & & 82.6\%& 82.1\%& 79.1\% \\
\hline
Digit 6& & & & 94.9\%& 95.0\% \\
\hline
Digit 7& & & & 92.9\%& 89.0\% \\
\hline
Digit 8& & & & & 85.5\% \\
\hline
Digit 9& & & & & 83.2\% \\
\hline
All& 99.9\%& 95.8\%& 93.0\%& 91.0\%& 87.3\% \\
\hline
\end{tabular}
\label{tab:mnistonline}
\end{center}
Results of simulated online learning with HD-Glue on the MNIST dataset. We gradually introduce new models, each of which handles discriminating between two digits. All models are presented with 100 training examples, for each existing class - old classes retain their training. We show the performance per each class, as well as the total across existing classes. By the time all models are introduced, HD-Glue has been trained on 3000 examples, unbalanced across classes. The overall performance is the same as if training HD-Glue on the full 3000 examples from the beginning. Note that \textbf{no} error correction has been done here, nor multiple models.
\end{table}

\subsection{CIFAR-10 Results}\label{cifar10res}

Results for the CIFAR-10 data-set are shown in Table~\ref{tab:cifar10}. Once again, we use simple image encoders to learn representations, which then predict the class using a final layer. We use the signal values of this output layer to either encode hypervector representations, or use the real valued vector of signals in benchmark algorithms. We keep the output layer the same number of components across all 20 encoders generated, each encoder with a different initialization. Results are split into 10 model consensus and 20, to show the improvement of HD-Glue with more models. We also use different dimension sizes for the hypervectors to show how HD-Glue improves with higher hyperdimensionality. In all but two cases, HD-Glue outperforms all benchmarks with 10 models, but outperforms all other models at 20, showing this improvement.

\begin{table*}[h]
\caption{HD-Glue performance vs. benchmark methods on the CIFAR-10 data-set}
\begin{center}
\begin{tabular}{|c|c|c|c|c|c|c|c|c|c|c|c|}
\hline
\multicolumn{12}{|c|}{\textbf{CIFAR-10 Data-set Benchmarking Results (Note: Single Neural Network Accuracy is 54.5\%)}} \\
\hline
\textbf{\#Models=10}&\multicolumn{3}{|c|}{\textbf{HD-Glue}}&\multicolumn{8}{|c|}{\textbf{Benchmark Algorithms}} \\
\hline
 \textbf{Size}& \textbf{dim=2000}& \textbf{dim=4000}& \textbf{dim=8000}& \textbf{\textit{KNN}}& \textbf{\textit{L-SVM}}& \textbf{\textit{RBF-SVM}}& \textbf{\textit{DT}}& \textbf{\textit{RF}}& \textbf{\textit{Log-Reg}}& \textbf{\textit{AdaBoost}}& \textbf{\textit{NB}}  \\
\hline
10& 52.0\%& 53.4\%& 54.6\% (-1.0\%)& 23.5\%& \textbf{55.6\%}& \textbf{55.6\%}& 23.4\%& 24.4\%& 54.6\%& 11.2\%& \textbf{55.6\%} \\
\hline
50& 60.4\%& 61.4\%& \textbf{62.3\% (+1.2\%)}& 58.5\%& 61.0\%& 35.0\%& 24.3\%& 39.3\%& 61.0\%& 17.1\%& 46.0\% \\
\hline
100& 61.5\%& 62.7\%& \textbf{63.7\% (+1.7\%)}& 60.5\%& 61.3\%& 34.5\%& 34.1\%& 41.1\%& 62.0\%& 10.5\%& 57.9\% \\
\hline
500& 64.1\%& 64.5\%& 65.1\% (-0.2\%)& 62.3\%& \textbf{65.3\%}& 24.3\%& 44.2\%& 49.2\%& 63.3\%& 28.0\%& 64.3\% \\
\hline
1000& 62.9\%& 64.4\%& \textbf{65.4\% (+0.1\%)}& 62.7\%& 65.3\%& 23.7\%& 42.6\%& 52.4\%& 63.0\%& 33.4\%& 64.7\% \\
\hline
\textbf{\#Models=20}& \multicolumn{3}{|c|}{\textbf{HD-Glue}}& \multicolumn{8}{|c|}{\textbf{Benchmark Algorithms}} \\
\hline
 \textbf{Size}& \textbf{dim=2000}& \textbf{dim=4000}& \textbf{dim=8000}& \textbf{\textit{KNN}}& \textbf{\textit{L-SVM}}& \textbf{\textit{RBF-SVM}}& \textbf{\textit{DT}}& \textbf{\textit{RF}}& \textbf{\textit{Log-Reg}}& \textbf{\textit{AdaBoost}}& \textbf{\textit{NB}}  \\
\hline
10& 52.6\%& \textbf{56.0\%}& \textbf{56.0\% (+0.4\%)}& 23.5\%& 55.6\%& 55.6\%& 18.7\%& 16.9\%& 54.4\%& 12.3\%& 55.6\% \\
\hline
50& 60.7\%& 62.5\%& \textbf{63.8\% (+2.8\%)}& 58.3\%& 61.0\%& 42.1\%& 29.0\%& 34.2\%& 60.9\%& 16.4\%& 53.4\% \\
\hline
100& 62.1\%& 64.1\%& \textbf{64.3\% (+0.4\%)}& 60.8\%& 63.6\%& 39.1\%& 29.4\%& 41.7\%& 63.9\%& 15.7\%& 60.1\% \\
\hline
500& 65.0\%& 65.8\%& \textbf{66.2\% (+0.2\%)}& 63.4\%& 66.0\%& 39.6\%& 39.7\%& 51.6\%& 64.6\%& 25.8\%& 65.5\% \\
\hline
1000& 64.9\%& \textbf{66.2\%}& \textbf{66.2\% (+0.0\%)}& 63.7\%& \textbf{66.2\%}& 34.5\%& 47.0\%& 54.5\%& 64.7\%& 41.8\%& 65.8\% \\
\hline
\end{tabular}
\label{tab:cifar10}
\end{center}
Results for CIFAR-10 are shown above on benchmarks. The size column is the number of training examples shown to the neural networks, the best of which performs at 54.5\% accuracy, with each example creating an embedding in the output layer of the networks. The algorithm names are the same as in Table~\ref{tab:mnist}, with NB being Naive Bayes and L-SVM being Linear SVM, shortened to conserve space, and Log-Reg being Logistic Regression. Our benchmark algorithms use the element-wise average of the real-valued embeddings. We present results for both using 10 different, image encoder networks and for 20, acting in consensus. We also present results for different dimension length of hypervectors. Bolded results are the best performing, and for our best performing HD-Glue model we show how much better it is than the second best algorithm in terms of percent, and how much worse it is than the best performing one if it is not the best. As we can see, HD-Glue is the best performing model overall, and is very close to the best when it is not. 
\end{table*}

Additionally, we include results for using stronger networks on CIFAR-10, namely VGG11 \cite{simonyan2014very}, in Table~\ref{tab:cifar10pre}. Here, we use high performing networks trained only on CIFAR-10 to assess how well consensus through HD-Glue works in this scenario. We have dropped poorly performing benchmarks here for simplicity. When the networks are too similar, the benefits of HD-Glue are not as apparent. HD-Glue prefers diverse networks in its consensus. Still, the results are competitive, while maintaining the benefits of HD representations. It should be noted that Logistic Regression likely performs so well in this experiment because of how similar the models are. However, Linear-SVM catches up to this with enough examples.

\begin{table*}[h]
\caption{HD-Glue performance vs. benchmark methods on the CIFAR-10 data-set with high-performance networks}
\begin{center}
\begin{tabular}{|c|c|c|c|c|c|c|c|}
\hline
\multicolumn{8}{|c|}{\textbf{CIFAR-10 Data-set Benchmarking Results (VGG11 \#1: 92.1\%, VGG11 \#2: 91.1\%, VGG11 \#3: 90.8\%)}} \\
\hline
 &\multicolumn{3}{|c|}{\textbf{HD-Glue}}&\multicolumn{4}{|c|}{\textbf{Benchmark Algorithms}} \\
\hline
 \textbf{Size}& \textbf{dim=4000}& \textbf{dim=8000}& \textbf{dim=12000}& \textbf{\textit{KNN}}& \textbf{\textit{Linear SVM}}& \textbf{\textit{Log-Reg}}& \textbf{\textit{Naive Bayes}} \\
\hline
10& 77.4\%& 80.0\%& \textbf{81.2\% (+0.7\%)}& 28.6\%& 80.5\%& 80.0\%& 80.5\% \\
\hline
50& 87.5\%& 88.8\% (-1.6\%)& 88.7\%& 87.1\%& 86.7\%& \textbf{90.4\%}& 86.3\% \\
\hline
100& 88.9\%& 89.1\%& 89.4\% (-0.6\%)& 88.4\%& 89.3\%& \textbf{91.0\%}& 88.8\% \\
\hline
500& 89.0\%& 90.8\%& 90.9\% (-0.6\%)& 90.3\%& 91.2\%& \textbf{91.6\%}& 90.3\% \\
\hline
1000& 89.6\%& 90.8\%& 90.9\% (-0.6\%)& 90.9\%& \textbf{91.6\%}& \textbf{91.6\%}& 90.5\% \\
\hline
\end{tabular}
\label{tab:cifar10pre}
\end{center}
Results for CIFAR-10 with trained VGG11 models. Much like in Table~\ref{tab:mnist} and Table~\ref{tab:cifar10}, we compare HD-Glue to the benchmark algorithms. We only include competitive benchmarks here that perform similar to HD-Glue. Once again, size here denotes the number of training examples presented to each algorithm. Networks utilized were 3 incarnations of VGG11 \cite{simonyan2014very} obtained by training on CIFAR-10 using different initializations. Their accuracies were 92.1\%, 91.1\%, and 90.8\%, respectively. Bolded entries are best performing, and our best performing HD-Glue model includes how much better it was than the second best algorithm, or how much worse it was than the best, in terms of percent.
\end{table*}

\subsection{CIFAR-100 Results}

In the case of CIFAR-100, we have an opportunity to challenge HD-Glue in a harder setting. The 100 classes of CIFAR-100 are easily handled as hypervectors can perfectly recall hundreds of aggregated vectors \cite{mitrokhin2019learning}. For obtuse numbers of classes, multiple hypervectors can be used to handle subsets of classes. In Table~\ref{tab:cifar100}, we show the results of our benchmarks using a diverse set of image classification models, trained on CIFAR-100. Namely, we use VGG11, VGG13, VGG16, and VGG19 from \cite{simonyan2014very}, and ResNet18 and ResNet34 from \cite{he2016deep}. Once again, HD-Glue outperforms all other benchmarks. Compared to the results in Table~\ref{tab:cifar10pre}, this implies diverse networks are preferable in consensus as opposed to the same network with different initializations.

Moreover, we also show the results of CIFAR-100 in the case of networks with differing embedding lengths in Table~\ref{tab:cifar100diffembeds}. We use the same networks, however ResNet18 and ResNet34 have been retrained with a lower number of output signals, by half. This lowers their performance a little. We can see that for reasonable hypervector lengths, HD-Glue consensus will outperform ResNet34, the best performing individual model, after just 500 training examples.

\begin{table*}[htp]
\caption{HD-Glue performance vs. benchmark methods on the CIFAR-100 data-set with diverse networks}
\begin{center}
\begin{tabular}{|c|c|c|c|c|c|c|c|c|}
\hline
\multicolumn{9}{|c|}{\textbf{CIFAR-100 Data-set Benchmarking Results}} \\
\hline
\textbf{Single Model}& \textbf{\textit{VGG11}}& \textbf{\textit{VGG13}}& \textbf{\textit{VGG16}}& \textbf{\textit{VGG19}}& \textbf{\textit{ResNet18}}& \textbf{\textit{ResNet34}}& \multicolumn{2}{|c|}{ } \\
\hline
\textbf{Accuracy}& 65.8\%& 67.8\%& 65.4\%& 60.3\%& 75.9\%& 76.9\%& \multicolumn{2}{|c|}{ } \\
\hline
\multicolumn{9}{|c|}{ } \\
\hline
 &\multicolumn{3}{|c|}{\textbf{HD-Glue}}&\multicolumn{5}{|c|}{\textbf{Benchmark Algorithms}} \\
\hline
 \textbf{Size}& \textbf{dim=4000}& \textbf{dim=8000}& \textbf{dim=12000}& \textbf{\textit{KNN}}& \textbf{\textit{Linear SVM}}& \textbf{\textit{RBF-SVM}}& \textbf{\textit{Neural Net}}& \textbf{\textit{Naive Bayes}} \\
\hline
100& 66.2\%& 71.2\%& \textbf{72.1\% (+7.1\%)}& 26.5\%& 65.0\%& 65.0\%& 57.7\%& 65.0\% \\
\hline
500& 71.5\%& 76.0\%& \textbf{77.5\% (+2.4\%)}& 72.7\%& 75.1\%& 39.2\%& 74.9\%& 66.7\% \\
\hline
1000& 75.9\%& 76.9\%& \textbf{77.6\% (+0.9\%)}& 74.1\%& 76.7\%& 30.8\%& 76.1\%& 75.8\% \\
\hline
5000& 76.8\%& 77.9\%& \textbf{78.0\% (+0.0\%)}& 76.1\%& \textbf{78.0\%}& 23.3\%& 76.8\%& 77.6\% \\
\hline
\end{tabular}
\label{tab:cifar100}
\end{center}
Results for CIFAR-100 with diverse image classification models. Much like in Table~\ref{tab:mnist}, Table~\ref{tab:cifar10}, and Table~\ref{tab:cifar10pre}, we compare HD-Glue to the benchmark algorithms. We only include competitive benchmarks here that perform similar to HD-Glue. Once again, size here denotes the number of training examples presented to each algorithm. Networks utilized were VGG11, VGG13, VGG16, VGG19 \cite{simonyan2014very}, ResNet18 and ResNet34 \cite{he2016deep}, obtained by training on CIFAR-100 using different initializations. Their accuracies are shown in the table. Bolded entries are best performing, and our best performing HD-Glue model includes how much better it was than the second best algorithm, or how much worse it was than the best, in terms of percent. As we can see, HD-Glue achieved the best accuracy across all sizes.
\end{table*}

\section{Discussion}\label{discussion}

In here we discuss the results in section \ref{results}, their implications and future work.

\subsection{Observations from Results}

\begin{table}[h]
\caption{HD-Glue with different embedding sizes}
\begin{center}
\begin{tabular}{|c|c|c|c|}
\hline
\multicolumn{4}{|c|}{\textbf{CIFAR-100 Different Embedding Sizes}} \\
\hline
 & \multicolumn{3}{|c|}{\textbf{Length}} \\
\hline
 & \multicolumn{2}{|c|}{\textbf{512}}& \multicolumn{1}{|c|}{\textbf{256}} \\
\hline
\textbf{Model}&
\textbf{\textit{VGG11}}&
\textbf{\textit{VGG13}}& \textbf{\textit{ResNet18}} \\
\hline
\textbf{Accuracy}& 65.8\%& 67.8\%& 72.3\% \\
\hline
\textbf{Model}&
\textbf{\textit{VGG16}}& \textbf{\textit{VGG19}}& \textbf{\textit{ResNet34}} \\
\hline
\textbf{Accuracy}& 65.4\%& 60.3\%& \textbf{74.5}\% \\
\hline
\multicolumn{4}{|c|}{ } \\
\hline
 &\multicolumn{3}{|c|}{\textbf{HD-Glue}} \\
\hline
 \textbf{Size}& \textbf{dim=4000}& \textbf{dim=8000}& \textbf{dim=12000} \\
\hline
100& 64.1\%& 68.4\%& 70.5 \\
\hline
500& 72.7\%& \textbf{74.8}\%& \textbf{75.2}\% \\
\hline
1000& 73.1\%& \textbf{75.5}\%& \textbf{75.9}\% \\
\hline
5000& 74.1\%& \textbf{76.2}\%& \textbf{\underline{76.3}}\% \\
\hline
\end{tabular}
\label{tab:cifar100diffembeds}
\end{center}
Results for CIFAR-100 with networks that have differing sizes. We present only the results of HD-Glue, as other benchmarking methods cannot account for different vector lengths. Specifically, the ResNet networks have half sized embeddings (which lowers performance a bit). We mark the performance of the best individual network in bold, and also mark HD-Glue models that outperform it in bold. In these instances, the HD-Glue consensus outperforms individual networks. We underline the best performing HD-Glue model. As we can see, for reasonable hypervector lengths, consensus performs better after a small number of examples.
\end{table}

It seems clear that HD-Glue best performs when very different networks are combined together, supporting the results in \cite{mitrokhin2020symbolic}. One key observation from Table~\ref{tab:cifar10pre}'s results, is that a limitation of our hyperdimensional consensus manifests for too similar networks. A simple question arises: can these results be improved? Consider the intersection of all correct answers each of the individual 3 VGG networks provide: a performance of 94.6\% can be achieved if the correct result is chosen from the 3 networks. When best 2 out of 3 consensus is considered, the accuracy falls to 90.6\%, comparable to our results in Table~\ref{tab:cifar10pre} for HD-Glue. Of that discrepancy, HD-Glue itself correctly classifies 0.58\% of examples that our intersected models cannot. That leaves an improvement of 4.5\% that is still within the realms of possibility - meaning HD-Glue can be further improved. However, it's clear that a more complicated HD structure than the HIL is necessary to lessen this gap.

Furthermore, as the networks are trained, HD-Glue very quickly approaches a higher performance, bootstrapping the consensus - this was also observed in \cite{mitrokhin2020symbolic}. However, HD-Glue has the benefit of opening the door to many other neural network architectures that can be glued together, due to using the output signals of the network directly. Moreover, it is competitive to other methods that can be used for quick consensus prediction, without needing end-to-end learning. In fact, with diverse networks in consensus, HD-Glue can outperform these methods. Given how fast and cheap computations in HD-Glue are, these properties are well-worth the effort, especially since similar real-valued vector methods would require consistent vector lengths, a restriction we do not have with hypervectors.

\subsection{Implications of Results}

Suppose the modalities of glued models in HD-Glue go beyond that of image domain. In theory, one can combine classical image models with spectral models, log-polar models, etc. In fact, there is nothing stopping the creation of a super-model that incorporates every pre-trained image model in existence; the only limitation here is how many neural models can be loaded in parallel. In a distributed setting, this idea becomes far more applicable. Furthermore, what if entire fields are bridged with gluing? Indeed, the idea of fusing video, audio, and text based models together poses an interesting idea; can such a consensus model decide on what sort of data is relevant to a task? These are open questions to future work.

The implication that more diverse networks increase the performance of glued models indicates a new potential way to tackle the problem of machine learning: \textit{targeting specific aspects of the task to solve with specifically constructed models that use resources in a multifaceted approach, with less resources used, that are glued together.} An interesting side effect of such an approach, is that transfer learning is much easier, since we could simply use the hypervectors as input to our new task, and constrain learning to that domain.

\subsection{Future Work}

One big limitation to HD-gluing is the fact that neural network models are needed to begin with. It is generally infeasible, outside of distributed settings, to have dozens of large neural networks operating in tandem at testing time. There is a pressing need for a more advanced learning strategy than compressing many neural models to a single hypervector. While what HD-Glue can achieve through a single hypervector model is impressive, there is clearly a need to use many hypervectors together, in order to extract more knowledge from the neural network to the hypervector. 

In such a construct, transfer learning into a purely hyperdimensional ``network" of hypervectors can omit the need to keep the neural network models around at all. If the student hypervector model accepts some kind of encoding of the input directly, then it can transfer learn what to do with that input to classify from the teacher neural network. After training, the teacher network can be thrown away and a purely hyperdimensional model remains. This would provide a hypercompression of neural models, which can be put directly into autonomous vehicles, using minimal space or resources.

Another big avenue of future work lies in the life-long-learning aspects of HD-gluing. In our results, we performed learning until 100\% performance on the training set was achieved, through the error handling method described in section \ref{multihypervectors}. But the testing set itself could also be incorporated into this strategy. In an online learning setting, HD-gluing permits forming models that can achieve 100\% accuracy on all examples seen so far, and form compact memories of previous examples that it can store to aid this process. Additionally, learned hypervector models that have been glued together can be compressed together into a further compact memory, to free up room for new models. This process can continue indefinitely. New classes could come into the picture, or new neural networks. What are the limitations of this as a life-long-learning strategy, and how does it compare to the state of the art? This is a question outside the scope of this paper.

\section{Conclusion}\label{conclusion}

In conclusion, we have presented a technique for gluing neural network models together at a symbolic level through hypervector representations of their output signals, using HC operations to then classify across all networks in consensus. This technique has life-long-learning applications, can learn online, and is highly adaptable and dynamic; tolerant to networks disappearing or new networks appearing to be added to the glued models. Furthermore, the architectures of glued networks need not be similar nor of the same modalities. This allows re-using existing networks to achieve higher performance through efficient and cheap consensus. We have shown this to be as powerful, if not more, than other standard methods of classification from signals directly, which do not have such properties. Finally, we discussed implications and potential uses of HD-Glue in future work, giving rise to interesting questions about what can be done with glued models.

\section{Acknowledgment}\label{acknowledgements}

The support of the Army Research Laboratory under a cooperative agreement
with the University of Maryland (ArtIAMAS project), and  NSF under grant OISE 2020624 are gratefully acknowledged.






\bibliographystyle{IEEEtran}
\bibliography{IJCNN-HD-Consensus-Paper}

\begin{thebibliography}{10}
\providecommand{\url}[1]{#1}
\csname url@samestyle\endcsname
\providecommand{\newblock}{\relax}
\providecommand{\bibinfo}[2]{#2}
\providecommand{\BIBentrySTDinterwordspacing}{\spaceskip=0pt\relax}
\providecommand{\BIBentryALTinterwordstretchfactor}{4}
\providecommand{\BIBentryALTinterwordspacing}{\spaceskip=\fontdimen2\font plus
\BIBentryALTinterwordstretchfactor\fontdimen3\font minus
  \fontdimen4\font\relax}
\providecommand{\BIBforeignlanguage}[2]{{%
\expandafter\ifx\csname l@#1\endcsname\relax
\typeout{** WARNING: IEEEtran.bst: No hyphenation pattern has been}%
\typeout{** loaded for the language `#1'. Using the pattern for}%
\typeout{** the default language instead.}%
\else
\language=\csname l@#1\endcsname
\fi
#2}}
\providecommand{\BIBdecl}{\relax}
\BIBdecl

\bibitem{kanerva2009hyperdimensional}
P.~Kanerva, ``Hyperdimensional computing: An introduction to computing in
  distributed representation with high-dimensional random vectors,''
  \emph{Cognitive computation}, vol.~1, no.~2, pp. 139--159, 2009.

\bibitem{mitrokhin2019learning}
A.~Mitrokhin, P.~Sutor, C.~Ferm{\"u}ller, and Y.~Aloimonos, ``Learning
  sensorimotor control with neuromorphic sensors: Toward hyperdimensional
  active perception,'' \emph{Science Robotics}, vol.~4, no.~30, 2019.

\bibitem{mitrokhin2020symbolic}
A.~Mitrokhin, P.~Sutor, D.~Summers-Stay, C.~Ferm{\"u}ller, and Y.~Aloimonos,
  ``Symbolic representation and learning with hyperdimensional computing,''
  \emph{Frontiers in Robotics and AI}, vol.~7, p.~63, 2020.

\bibitem{widdows2015reasoning}
D.~Widdows and T.~Cohen, ``Reasoning with vectors: A continuous model for fast
  robust inference,'' \emph{Logic Journal of the IGPL}, vol.~23, no.~2, pp.
  141--173, 2015.

\bibitem{rahimi2016hyperdimensional}
A.~Rahimi, S.~Benatti, P.~Kanerva, L.~Benini, and J.~M. Rabaey,
  ``Hyperdimensional biosignal processing: A case study for emg-based hand
  gesture recognition,'' in \emph{2016 IEEE International Conference on
  Rebooting Computing (ICRC)}.\hskip 1em plus 0.5em minus 0.4em\relax IEEE,
  2016, pp. 1--8.

\bibitem{rahimi2018efficient}
A.~Rahimi, P.~Kanerva, L.~Benini, and J.~M. Rabaey, ``Efficient biosignal
  processing using hyperdimensional computing: Network templates for combined
  learning and classification of exg signals,'' \emph{Proceedings of the IEEE},
  vol. 107, no.~1, pp. 123--143, 2018.

\bibitem{rachkovskiy2005sparse}
D.~A. Rachkovskiy, S.~V. Slipchenko, I.~S. Misuno, E.~M. Kussul, and T.~N.
  Baidyk, ``Sparse binary distributed encoding of numeric vectors,''
  \emph{Journal of Automation and Information Sciences}, vol.~37, no.~11, 2005.

\bibitem{sutor2018computational}
P.~Sutor, D.~Summers-Stay, and Y.~Aloimonos, ``A computational theory for
  life-long learning of semantics,'' in \emph{International Conference on
  Artificial General Intelligence}.\hskip 1em plus 0.5em minus 0.4em\relax
  Springer, 2018, pp. 217--226.

\bibitem{benediktsson1997classification}
J.~A. Benediktsson and J.~R. Sveinsson, ``Classification of hyperdimensional
  data using data fusion approaches,'' in \emph{IGARSS'97. 1997 IEEE
  International Geoscience and Remote Sensing Symposium Proceedings. Remote
  Sensing-A Scientific Vision for Sustainable Development}, vol.~4.\hskip 1em
  plus 0.5em minus 0.4em\relax IEEE, 1997, pp. 1669--1671.

\bibitem{benediktsson1999classification}
J.~A. Benediktsson and I.~Kanellopoulos, ``Classification of multisource and
  hyperspectral data based on decision fusion,'' \emph{IEEE Transactions on
  Geoscience and Remote Sensing}, vol.~37, no.~3, pp. 1367--1377, 1999.

\bibitem{cheung2019superposition}
B.~Cheung, A.~Terekhov, Y.~Chen, P.~Agrawal, and B.~Olshausen, ``Superposition
  of many models into one,'' \emph{Advances in Neural Information Processing
  Systems}, vol.~32, pp. 10\,868--10\,877, 2019.

\bibitem{kleyko2018classification}
D.~Kleyko, A.~Rahimi, D.~A. Rachkovskij, E.~Osipov, and J.~M. Rabaey,
  ``Classification and recall with binary hyperdimensional computing: Tradeoffs
  in choice of density and mapping characteristics,'' \emph{IEEE transactions
  on neural networks and learning systems}, vol.~29, no.~12, pp. 5880--5898,
  2018.

\bibitem{sagi2018ensemble}
O.~Sagi and L.~Rokach, ``Ensemble learning: A survey,'' \emph{Wiley
  Interdisciplinary Reviews: Data Mining and Knowledge Discovery}, vol.~8,
  no.~4, p. e1249, 2018.

\bibitem{dietterich2002ensemble}
T.~G. Dietterich \emph{et~al.}, ``Ensemble learning,'' \emph{The handbook of
  brain theory and neural networks}, vol.~2, no.~1, pp. 110--125, 2002.

\bibitem{polikar2012ensemble}
R.~Polikar, ``Ensemble learning,'' in \emph{Ensemble machine learning}.\hskip
  1em plus 0.5em minus 0.4em\relax Springer, 2012, pp. 1--34.

\bibitem{dong2020survey}
X.~Dong, Z.~Yu, W.~Cao, Y.~Shi, and Q.~Ma, ``A survey on ensemble learning,''
  \emph{Frontiers of Computer Science}, vol.~14, no.~2, pp. 241--258, 2020.

\bibitem{joshi2016language}
A.~Joshi, J.~T. Halseth, and P.~Kanerva, ``Language geometry using random
  indexing,'' in \emph{International Symposium on Quantum Interaction}.\hskip
  1em plus 0.5em minus 0.4em\relax Springer, 2016, pp. 265--274.

\bibitem{imani2019quanthd}
M.~Imani, S.~Bosch, S.~Datta, S.~Ramakrishna, S.~Salamat, J.~M. Rabaey, and
  T.~Rosing, ``Quanthd: A quantization framework for hyperdimensional
  computing,'' \emph{IEEE Transactions on Computer-Aided Design of Integrated
  Circuits and Systems}, vol.~39, no.~10, pp. 2268--2278, 2019.

\bibitem{deng2012mnist}
L.~Deng, ``The mnist database of handwritten digit images for machine learning
  research,'' \emph{IEEE Signal Processing Magazine}, vol.~29, no.~6, pp.
  141--142, 2012.

\bibitem{krizhevsky2009learning}
A.~Krizhevsky \emph{et~al.}, ``Learning multiple layers of features from tiny
  images,'' 2009.

\bibitem{simonyan2014very}
K.~Simonyan and A.~Zisserman, ``Very deep convolutional networks for
  large-scale image recognition,'' \emph{arXiv preprint arXiv:1409.1556}, 2014.

\bibitem{he2016deep}
K.~He, X.~Zhang, S.~Ren, and J.~Sun, ``Deep residual learning for image
  recognition,'' in \emph{Proceedings of the IEEE conference on computer vision
  and pattern recognition}, 2016, pp. 770--778.

\bibitem{fix1989discriminatory}
E.~Fix and J.~L. Hodges, ``Discriminatory analysis. nonparametric
  discrimination: Consistency properties,'' \emph{International Statistical
  Review/Revue Internationale de Statistique}, vol.~57, no.~3, pp. 238--247,
  1989.

\bibitem{cortes1995support}
C.~Cortes and V.~Vapnik, ``Support-vector networks,'' \emph{Machine learning},
  vol.~20, no.~3, pp. 273--297, 1995.

\bibitem{quinlan1986induction}
J.~R. Quinlan, ``Induction of decision trees,'' \emph{Machine learning},
  vol.~1, no.~1, pp. 81--106, 1986.

\bibitem{ho1995random}
T.~K. Ho, ``Random decision forests,'' in \emph{Proceedings of 3rd
  international conference on document analysis and recognition}, vol.~1.\hskip
  1em plus 0.5em minus 0.4em\relax IEEE, 1995, pp. 278--282.

\bibitem{freund1997decision}
Y.~Freund and R.~E. Schapire, ``A decision-theoretic generalization of on-line
  learning and an application to boosting,'' \emph{Journal of computer and
  system sciences}, vol.~55, no.~1, pp. 119--139, 1997.

\bibitem{rish2001empirical}
I.~Rish \emph{et~al.}, ``An empirical study of the naive bayes classifier,'' in
  \emph{IJCAI 2001 workshop on empirical methods in artificial intelligence},
  vol.~3, no.~22, 2001, pp. 41--46.

\end{thebibliography}


\end{document}